\documentclass[11pt,a4paper]{article}
\usepackage{latexsym}
\usepackage{graphicx}
\usepackage[all]{xypic}
\usepackage{epsf}
\usepackage{color}
\usepackage{algorithm}
\usepackage{algpseudocode}

\usepackage{amssymb}

\newtheorem{theorem}{Theorem}
\newtheorem{proposition}[theorem]{Proposition}
\newtheorem{lemma}[theorem]{Lemma}
\newtheorem{corollary}[theorem]{Corollary}
\newtheorem{definition}{Definition}
\newtheorem{remark}{Remark}

\newenvironment{proof}{\noindent\textit{Proof. \  }}{\hfill$\Box$\medskip\par}
\newenvironment{proofOf}[1]{\noindent\textbf{Proof of {#1}.}}{\hfill$\Box$\medskip\par}

\newcommand{\enumpb}[1]{\textsc{Enum}\!\cdot\!{#1}}
\newcommand{\constantdelaylin}{\textsc{Constant-Delay$_{lin}$}}
\newcommand{\constantdelaypoly}{\textsc{Constant-Delay$_{poly}$}}
\newcommand{\enumAlgo}[1]{\textsc{enum}({#1})}
\newcommand{\precompAlgo}[1]{\textsc{precomp}({#1})}
\newcommand{\delayAlgo}[1]{\textsc{delay}({#1})}

\renewcommand{\emph}[1]{\textsl{#1}}

\newcommand{\calA}{{\cal A}}
\newcommand{\calB}{{\cal B}}

\newcommand{\calS}{{\cal S}}

\newcommand{\st}[2]{\ensuremath{\left\langle \mathit{#1} ; #2 \right\rangle}}
\newcommand{\su}[2]{\ensuremath{#1_{1}, \ldots, #1_{#2}}}

\newcommand{\raus}[1]{}

\newcommand{\DLIN}{\mbox{\rm DLIN}}

\def\fobij{\mathbf{FO_{Bij}}}
\def\foBoundDeg{\mathbf{FO_{Deg}}}

\newcommand{\sigmastructure}{\sigma\mbox{-structure}}
\newcommand{\sigmastructures}{\sigma\mbox{-structures}}

\newcommand{\barre}[1]{\overline{#1}}

\newcommand {\ex}{\exists}
\newcommand {\exk}[2]{\exists^{#1}_{#2}}

\newcommand {\sg}{\sigma}
\renewcommand{\t}{\tau}
\newcommand {\eps}{\epsilon}

\def\tu#1{\overline{#1}}

\newcommand{\modelchecking}[1]{\textsc{MC}({#1})}
\newcommand{\query}[1]{\textsc{Query}({#1})}

\newcounter{cpteur}

\begin{document}

\fontencoding{OT1}
 
\fontfamily{ppl}

\title{First-order queries on structures of bounded degree are computable with constant delay}

\author{
Arnaud Durand~\thanks{
  LACL - CNRS FRE 2673,~D\'epartement d'informatique,
  Universit{\'e} Paris~12,  94010 Cr{\'e}teil - France.
 Email: \texttt{durand@univ-paris12.fr}}
  \and Etienne Grandjean~\thanks{
GREYC - CNRS UMR 6072, Universit\'e de Caen - Campus 2, F-14032
Caen  cedex - France. Email: {\tt grandjean@info.unicaen.fr}}
}

\maketitle

\begin{abstract}
A bounded degree structure is either a relational structure all of whose relations  are of bounded degree or a functional structure involving bijective functions only. In this paper, we   revisit the complexity of the evaluation problem of not necessarily Boolean first-order queries over structures of bounded degree. Query evaluation is considered here as a dynamical process. We  prove that any query on bounded degree structures is $\constantdelaylin$, i.e.,  can be computed by an algorithm that has two separate parts: it has a precomputation step of linear time in the size of the structure and then, it outputs all tuples one by one with a constant (i.e. depending on the size of the formula only) delay between each.  Seen as a global process, this implies that queries on bounded structures can be evaluated in total time $O(f(|\varphi|).(|\calS|+|\varphi(\calS)|))$ and space $O(f(|\varphi|).|\calS|)$ where $\calS$ is the structure, $\varphi$ is the formula, $\varphi(\calS)$ is the result of the query and $f$ is some function.

Among other things, our results generalize a result of~\cite{Seese-96} on the data complexity of the model-checking problem for bounded degree structures. Besides, the originality of our approach compared to that~\cite{Seese-96} and comparable results  is that it does not rely on the  Hanf's model-theoretic  technic (see~\cite{Hanf-65}) and is completely effective. 
\end{abstract}

\section*{Introduction}

Evaluating the expressive power of logical formalisms is an important task in theoretical computer science. It has many applications in numerous fields such as complexity theory, verification or databases.
In this latter case, it often amounts to determine  
how difficult it is to compute a query written in a given language. 
In this vein, determining which fragments of first-order logic defines tractable query languages has deserved much attention.

It is well known, that 
over an arbitrary signature, computing a first-order query can be done in time polynomial in the size of the structure (and even in logarithmic space and $AC^0$). However the exponent of this polynomial depends heavily on the formula size (more precisely, on the number of variables). Nevertheless, for particular kinds of structures or formulas the complexity bound can be substantially improved. In~\cite{Seese-96}, it is proved that checking if a given first-order  sentence $\varphi$ is true (i.e., the Boolean query or model-checking problem) in a structure $\calS$ all of whose relations are of bounded degree  can be done in linear time in the size of $\calS$. The method used to prove this result relies on old model-theoretic technics (see~\cite{Hanf-65}). It is perfectly constructive but hardly implementable. Later, still using such kind of methods, several other tractability results have been shown for the complexity of the model-checking of first-order formulas over  structures or formulas that admit nice  (tree) decomposition properties (see~\cite{FlumFG-02}).

In this paper, a \textit{bounded degree structure} is either a relational structure all of whose relations  are of bounded degree or a functional structure involving bijective functions only.

The main goal of this paper is to  revisit the complexity of the evaluation problem of not necessarily Boolean first-order queries over structures of bounded degree.
We regard query evaluation as a \textit{dynamical process}. Instead of considering the cost of the evaluation globally, we measure the delay between consecutive tuples, i.e., query problems are viewed as enumeration problems. This latter kind of problems appears widely in many areas of computer science (see for example~\cite{EiterG-95,EiterGM-03,BorosGKM-00,KavvadiasSS-00,Goldberg-94} or~\cite{JohnsonYP-88} for basic complexity notions on enumeration). However, to our knowledge, relation to query evaluation has not been investigated so far.

We  prove that any query on bounded degree structures is $\constantdelaylin$, i.e.,  can be computed by an algorithm that has two separate parts: it has a precomputation step whose time complexity is linear in the size of the structure and then,  outputs all the solution tuples one by one with a constant (i.e., depending on the size of the formula only) delay between two successive tuples.  Seen as a global process, this implies that queries on bounded structures can be evaluated in total time $O(f(|\varphi|).(|\calS|+|\varphi(\calS)|))$ and space $O(f(|\varphi|).|\calS|)$ where $|\calS|$ is the size of the structure $\calS$, $|\varphi|$ is that of the formula $\varphi$, $|\varphi(\calS)|$ is the size of the result $\varphi(\calS)$ of the query and $f$ is some function. 
As a corollary,  it implies that the time complexity of the model-checking problem is $O(f(|\varphi|).|\calS|)$ thus providing an alternative proof of the result of~\cite{Seese-96}.

A particularity of the main method used in this paper is that it does not rely on model-theoretic technic as previous results of the same kind (see, for example,~\cite{Seese-96} or~\cite{Lindell-04} for a generalization to least-fixed point formulas). Instead, we develop a quantifier elimination method suitable for bijective unary functions and apply it to obtain our complexity bound. An advantage of this method is that it is effective and easily implementable.
Another advantage is that our paper is completely self-contained.

Besides, the $\constantdelaylin$ class is an interesting notion by itself and is, to our knowledge, a new complexity class for enumeration problems: as proved for linear time complexity (the class $\DLIN$ studied in~\cite{GrandjeanS-02}) it can be shown that $\constantdelaylin$ is a robust class and is in some sense the minimal  robust complexity class of enumeration problems. 

The paper is organized as follows. First, basic definitions are given in Section~\ref{definitions}. In particular, in Subsection~\ref{definition constant delay}, we recall definitions about enumeration problems and introduce the notion of constant delay computation and prove some basic properties about it. In Section~\ref{First-order queries on bijective structures}, the quantifier elimination method is introduced and is applied to the evaluation problem of first-order formulas over functional structures all of whose functions are bijective.
 In Section~\ref{SEC degre borne},
 using classical logical interpretation technics, this later problem is reduced  in linear time to the first-order query problem over structures of bounded degree thus providing the same bound for it. Finally, in Subsection~\ref{Complexity of subgraphs problems}, consequences about the complexity of the subgraph (resp. induced subgraph) isomorphism problem are given.

\section{Definitions}~\label{definitions}

\subsection{Logical definitions and query problems}
 We suppose the reader to be familiar with basic notions of first-order
logic.
A \textit{signature} $\sg$ is a finite set of relational and functional symbols of given arities ($0$-ary function symbols are constants symbols). The arity of $\sg$ is the maximal arity of its symbols. The set $\sg$ is called \textit{unary functional} if all its symbols are of arity bounded by one.

 A (finite)  $\sigmastructure$ consists of a domain $D$ together with an interpretation of each symbol of $\sg$ over $D$ (the same notation is used here for each signature symbol and its interpretation).

In this paper, we will distinguish between two kinds of signatures on which  semantical restrictions on their possible interpretation are imposed:

\begin{itemize}

\item Either $\sg$ is made of constant and monadic (i.e., unary) relation symbols and unary function symbols  whose interpretation is taken among bijective functions (i.e., permutations) only,

\item Or $\sg$ contains relation symbols only whose degrees are bounded by some given constant (detailed definitions about bounded degree relations are delayed till section~\ref{SEC degre borne}). 

\end{itemize}  
 
Structures defined by either of semantical restrictions will be called \textit{bounded degree structures}.
 
 In what follows we make precise notions and problems about first-order logic over bijective structures. 

\begin{definition}
Let $\sg = \{ \tu c, \tu U, \su{f}{k} \}$ be a signature consisting of constant symbols $c_i\in \tu c$, of monadic predicates $U_i\in \tu U$ and of unary
function symbols $f_i$, $i=1,\dots,k$. A {\em bijective $\sigmastructure$} is
a $\sigmastructure$ $\calS$ of the form $\calS =
\st{D}{\tu c, \tu U, \su{f}{k}}$ where each $f_i$ is a permutation on domain
$D$.
\end{definition}

One of the main results of this paper provides a quantifier elimination
method over bijective structures. As it is usual for such kind of
result, the elimination will be done in a richer language. The following definition is required.

\begin{definition}
A {\em bijective term} $\t (x)$ is of the form
$f_1^{\eps_1}\dots f_l^{\eps_l}(x)$ where $l\geq 0$, $x$ is a
variable and where each $f_i^{\eps_i}$ is either the function
symbol $f_i$  or its reciprocal $f_i^{-1}$. The term $\t^{-1}(x)$
denotes the reciprocal of the term $\t(x)$.

A {\em bijective atomic formula} is of one of the following four forms where $\t (x)$ and $\t_1 (x)$ are bijective terms:
\begin{itemize}
\item either a \textit{bijective equality} $\t (x) = \t_1 (y)$,

\item or $\t (x) = c$ where $c$ is a constant symbol,

\item or  $U(\t (x))$ where $U$ is a monadic predicate,

\item or a \textit{cardinality statement} $\exk{k}{x}\Psi(x)$
where the quantifier $\exk{k}{x}$ is interpreted as "there exist
at least $k$ values of $x$ such that" and $\Psi$ is a Boolean
combination of bijective atoms $\alpha (x)$ over variable $x$ only.
\end{itemize}
\end{definition}

As the reciprocal of each function symbol can be used, each
bijective equality $\t (x) = \t_1 (y)$ can be rephrased as
$\t_2 (x) = y$ where $\t_2 (x)=\t_1^{-1}\t(x)$. A {\em bijective literal} is a bijective atomic formula or its negation.

\begin{definition}
The set $\fobij$ of \text{bijective first-order formulas}  is the set of first-order formulas built over
bijective atomic formulas of some unary signature $\sg$.
\end{definition}

Let $\barre{t}=(t_1,\dots,t_k)$ be a $k$-tuple of variables and
 $\varphi(\barre{t})$ and $\varphi'(\barre{t})$ be two
$\sg$-formulas with free variables $\barre{t}$. Formulas
$\varphi(\barre{t})$ and $\varphi'(\barre{t})$ are
\textit{equivalent} if for all $\sigmastructures$ $\calS$ and all
tuples $\barre{a}$ of element of the domain with
$|\barre{a}|=|\barre{t}|$ it holds that:

\[
(\calS, \barre{a}) \models \varphi(\barre{t}) \mbox{ iff } (\calS,
\barre{a}) \models \varphi'(\barre{t}).
\]

In this paper query problems are considered for specific classes of first-order formulas (and structures).
One of the specific problems under consideration here is the following.

\medskip

\noindent $\query{\fobij}$\\
\noindent \textbf{Input:} a unary functional signature $\sg$, a bijective $\sg$-structure $\calS$ and a first-order bijective $\sg$-formula $\varphi(\tu x)$ with $k$ free variables $\tu x = (x_1,\dots,x_k)$\\  
\noindent \textit{Parameter:} $\varphi$ \\
\noindent \textbf{Output:} $\varphi(\calS) = \{\tu a \in D^k : (\calS, \tu a) \models \varphi(\tu x)\}$.

\medskip
The Boolean query problem (the subproblem where $k=0$) is often called a model-checking problem. It will be denoted by $\modelchecking{\fobij}$ here. 
As suggested by the formulation of the query problem, we are interested in its parameterized complexity and  the complexity results  given here  consider the size of the query formula $\varphi$ as the parameter (see~\cite{DowneyF-99}).

\subsection{Model of computation and measure of time}

The model of computation used in this paper is the Random Access
Machine (RAM) with uniform cost measure (see~\cite{AhoHU-74,
GrandjeanS-02, GrandjeanO-04, FlumFG-02}). As query problems are the main subject of this paper, instances of problems always consist of two kinds of objects: first-order structures and first-order  formulas.

The \textit{size} $|I|$ of an object $I$ is the number of
registers used to store $I$ in the RAM. If $E$ is the set $[n]$,
$|E|=card(E)=n$. If $R\subseteq D^k$ is a $k$-ary relation over
domain $D$, with $|D|=card(D)$, then $|R|=k.card(R)$: all the
tuples $(x_1,\dots,x_k)$ for which $R(x_1,\dots,x_k)$ holds must
be stored, each in a separate $k$-tuple of registers. Similarly, if $f$ is
a unary function from $D$ to $D$, all values
$f(x)$ must be stored and $|f|=|D|$.

 If $\varphi$ is a first-order formula, $|\varphi|$ is the number of
occurrences of variables, relation or function symbols and
syntactic symbols: $\exists, \forall, \wedge, \vee, \neg, =,  "(",
")", ","$. For example, if $\varphi \equiv \exists x \exists y \
R(x,y) \wedge \neg (x = y)$ then $|\varphi|=17$.

\bigskip

All the problems we consider in this paper are parameterized
problems: they take as input a list of objects made of a
$\sigma$-structure $\calS$ and a formula $\varphi$ and as output the result of the query size $\varphi(\calS)$. Due to the much larger size, in practice, of the structure $\calS$ than the size of formula $\varphi$, $|\calS|>>|\varphi|$, this latter one, $|\varphi|$ , in considered here as the parameter.

A problem \textbf{P} is said to be computable in time
$f(|\varphi|).T(|\calS|,|\varphi(\calS)|)$ for some function $f: N \rightarrow R^+$ if
there exists a RAM that computes \textbf{P} in time (i.e., the number
of instructions performed)  bounded by $f(|\varphi|).T(|\calS|,|\varphi(\calS)|)$ using
space,  i.e., addresses and register contents also bounded by
$f(|\varphi|).T(|\calS|,|\varphi(\calS)|)$.
The notation
$O_{\varphi}(T(|\calS|,|\varphi(\calS)|))$ is used when one does not want to make precise
the value of function $f$.  It is also assumed that the function
$T$ is at least linear and at most polynomial, i.e., $T(n,p) =
\Omega (n+p)$ and $T(n,p) = (n+p)^{O(1)}$. To give an example and to relate our complexity measure to the logarithmic cost measure, in case $T$ is linear, i.e., $T(n,p)=n+p$, the number of bits manipulated by the RAM is well linear in the number of bits needed to encode  the input and the output.

\subsection{Enumeration algorithms and constant delay computation}~\label{definition constant delay}

In this section, $A$ is a binary predicate. Enumeration problems will be defined by reference to such a predicate. 

\begin{definition}
Given a binary relation $A$, the enumeration function $\enumpb{A}$ associated to $A$ is defined as follows. For each input $x$:

\[
\enumpb{A(x)} = \{y \ : \ A(x,y) \mbox{ holds } \}
\]
\end{definition}

\begin{remark}
Query problems may evidently be seen as enumeration problems. The input $x$ is made of the structure $\calS$ and the formula $\varphi (\tu x)$, a witness $y$ is a tuple $\tu a$ and evaluating predicate $A$ amounts to check whether $(\calS,\tu a) \models \varphi(\tu x)$.    
\end{remark}

One may consider the delay between two consecutive solutions as an important point in the complexity of enumeration problems. In~\cite{JohnsonYP-88} several complexity measures for enumeration have been defined. One of the most interesting is that of \textit{polynomial delay} algorithm. An algorithm $\calA$ is said to run within a \textit{polynomial delay} if there is no more than a (fixed) polynomial delay between two consecutive solutions it outputs (and no more than a polynomial delay to output the first solution and between the last solution and the end of the algorithm). \textit{Polynomial delay} is often considered as the right notion of feasability for enumeration problems.

 In this paper, we introduce a much stronger complexity measure that forces  \textit{constant delay} between outputs.

\begin{definition}
An enumeration problem $\enumpb{A}$ is \textit{constant delay with linear precomputation}, which is written $\enumpb{A}\in \constantdelaylin$, if there exists a RAM algorithm $\calA$ which, for any input $x$, enumerates all the elements of the set $\enumpb{A(x)}$ with a constant delay, i.e., that satisfies the following  properties.  

\begin{enumerate}

\item $\calA$ uses linear input space, i.e., space $O(|x|)$

\item $\calA$ can be decomposed into the two following successive steps

\begin{enumerate}

\item $\precompAlgo{\calA}$ which runs some precomputations in time $O(|x|)$, and

\item $\enumAlgo{\calA}$ which outputs all solutions within a delay bounded by some constant $\delayAlgo{\calA}$. This delay applies between two consecutive solutions and after the last one.
\end{enumerate}

\end{enumerate}
\end{definition}

Allowing polynomial time precomputations (and polynomial space) instead of linear time, one may define a larger class called $\constantdelaypoly$.

\begin{remark}
As proved for the linear time class $\DLIN$ (see~\cite{GrandjeanS-02}), it can be shown that the complexity enumeration class $\constantdelaylin$ is robust, i.e., is not modified if the set of allowed operations and statements of the RAMs is changed in many ways. This is because linear time (and linear space) precomputations give the ability to precompute the 
tables of new allowed operations.
\end{remark}

The following result is immediate, it evaluates the total time cost of any constant delay algorithm.

\begin{lemma}~\label{LEM total time}
Let $\enumpb{A}$ be an enumeration problem belonging to $\constantdelaylin$ then, for any input $x$, the set $\enumpb{A(x)}$ can be computed in $O(|x| + |\enumpb{A(x)}|)$ total time, i.e., in time linear in the size of $|Input|+|Output|$, and linear input space $O(|x|)$.   
\end{lemma}

\begin{remark}
In the query problem we consider, the size of $\varphi$ is considered as a parameter. Then, $|x|=|\calS|$ and the constant delay depends on $|\varphi|$ only.
\end{remark}

The two lemmas below give basic properties of constant delay computations.

\begin{lemma}~\label{LEM linear time = constant delay}
An enumeration problem $\enumpb{A}$ computable in linear time $O(|x|)$  for any input $x$ belongs to $\constantdelaylin$. 
\end{lemma}

\begin{proof}
For any input $x$, one only has to compute the set $\enumpb{A(x)}$, to sort it and to eliminate the possible multiple occurrences of  solutions. These steps can be viewed as the precomputation part of the algorithm running in time $O(|x|)$. Then, one has to enumerate one by one the solutions of the  sorted list. This is obviously a constant delay process.
\end{proof}

\begin{lemma}~\label{LEM disjoint union}
Let $\enumpb{A}$ and $\enumpb{B}$ be two disjoint enumeration problems, i.e.,  such that, for any input  $x$, $\enumpb{A(x)} \cap \enumpb{B(x)} = \emptyset$. Let $\enumpb{(A\cup B)}$ be the union of this two enumeration problems defined by, for any $x$:

\[
\enumpb{(A\cup B)(x)} = \{y \ : \ A(x,y) \mbox{ or } B(x,y) \mbox{ holds }\}.
\]

If $\enumpb{A}$ and $\enumpb{B}$ belong to $\constantdelaylin$ then, problem $\enumpb{A\cup B}$
also belongs to $\constantdelaylin$.
\end{lemma}

\begin{proof}
Due to the disjointness of the two solutions sets for any input, the proof is evident. Given $\calA$ and $\calB$ the algorithms for problems $\enumpb{A}$ and $\enumpb{B}$, the following algorithm correctly computes for the problem $\enumpb{A\cup B}$.

\begin{algorithm}
\caption{Constant delay algorithm for $\enumpb{A\cup B}$}
\begin{algorithmic}[1]
\State \textbf{Input:} $x$
\State $\precompAlgo{\calA}$; $\precompAlgo{\calB}$
\State $\enumAlgo{\calA}$; $\enumAlgo{\calB}$ 
\end{algorithmic}
\end{algorithm}

Obviously, the delay is bounded by the maximum of $\delayAlgo{\calA}$ and $\delayAlgo{\calB}$.
\end{proof}

\begin{remark}
Note that the disjointness condition in the Lemma above is not always necessary. In case  there exist a total ordering $\leq$ and constant delay enumeration algorithms for  $\enumpb{A}$ and  $\enumpb{B}$ that enumerate solutions with respect to this unique ordering $\leq$ then, it is easily seen that    $\enumpb{A\cup B}$ belongs also to $\constantdelaylin$ even if the problems are not disjoints.
\end{remark}

\section{First-order queries on bijective structures}~\label{First-order queries on bijective structures}

\subsection{Quantifier elimination on bijective structures}

The key result of this paper consists of a quantifier elimination
method for $\fobij$ formulas.

\begin{theorem}[quantifier elimination for
$\fobij$]~\label{TH elimination quantificateur} Each bijective \textit{first-order}
formula is equivalent to a Boolean combination of bijective
\textit{atomic} formulas. More precisely, 
let $\varphi(\barre{t})\in \fobij$ with free variables
$\barre{t}$ then, there exists  a Boolean combination of bijective
\textit{atomic} formulas $\varphi'(\barre{t})$ over the same free
variables $\barre{t}$ equivalent  to $\varphi(\barre{t})$.

In the special case where $\varphi$ is closed (i.e., without free variable) then,  $\varphi$ is equivalent to a Boolean combination
of cardinality statements.

\end{theorem}

\begin{proof}
  As $\forall x \varphi \equiv \neg (\ex x \neg
\varphi)$, we only have to consider elimination of existentially
quantified variables. W.l.o.g., we consider formulas in disjunctive
normal form and, as existential quantifier commutes with
disjunction we may consider the case of the elimination of a
single existentially quantified variable $y$ in a formula of the
form:

\begin{equation}
\varphi(\barre{x})\equiv \ex y \ (\alpha_1 \wedge \dots \wedge
\alpha_r)
\end{equation}

\noindent where each $\alpha_i$ is a bijective literal
among variables $\barre{x}$ and $y$. Literals depending on
$\barre{x}$ only and cardinality statements need  not be
considered since they do not involve $y$, so
$\varphi(\barre{x})$ may be supposed of the following form:

\begin{equation}
\varphi(\barre{x})\equiv \ex y \ [\psi(y) \wedge y
=_{\eps_1}\t_1(x_{i_1}) \wedge \dots \wedge y
=_{\eps_k}\t_k(x_{i_k}) ]
\end{equation}

\noindent where each $y =_{\eps_j}\t_j(x_{i_j})$ with
$\eps_{j}=\pm 1$ is $y = \t_j(x_{i_j})$ if  $\eps_{j}= 1$ or  $y
\neq \t_j(x_{i_j})$ if  $\eps_{j}= -1$. To eliminate quantified
variable $y$ two cases may happen.

Suppose first there is at least one index $j$ such that $\eps_{j}=
1$. In this case, the equality $y = \t_j(x_{i_j})$ is used to
replace each occurrence of $y$ in the formula by the term
$\t_j(x_{i_j})$. The process results in a new formula
$\varphi'(\barre{x})$ without variable $y$.

The second possibility leads to a  more complicated replacement
scheme. Suppose that for every $j$, $\eps_j = -1$. Then,

\begin{equation}~\label{EQU cas 2}
\varphi(\barre{x})\equiv \ex y \ [\psi(y) \wedge \bigwedge_{j\leq
k} y\neq \t_j(x_{j}) ]
\end{equation}

(For simplicity of notations but w.l.o.g. we have supposed that $i_j
=j$ for $j=1,\dots,k$). The basic idea is now the following :
suppose $h\leq k$ is the number of distinct values among the $k$
terms  $\t_j(x_{j})$ such that $\psi(\t_j(x_{j}))$ is true; then,
formula $\varphi(\barre{x})$ is true if and only if the number of
$y$ such that $\psi(y)$ holds is strictly greater than $h$ (i.e.,
$\exk{h+1}{y}\psi(y)$ is true). Introducing (new) cardinality
statements in the formula, $\varphi(\barre{x})$ can be equivalently rephrased as the following Boolean combination of bijective atomic formulas:

\begin{equation}~\label{EQU cas 2 bis}
\varphi(\barre{x})\equiv
  \begin{array}[t]{l}
     {\displaystyle \bigvee_{h=0}^k   \bigvee_{P\subseteq [k], Q\subseteq P, |Q|=h}}\\
      \left[{\displaystyle \bigwedge_{j\in Q} \psi(\t_j(x_{j})) \wedge
    \bigwedge_{i\in P}\bigvee_{j\in Q} \t_i(x_{i})= \t_j(x_{j}) \wedge
    \bigwedge_{j\in [k]\setminus P}  \neg \psi(\t_j(x_{j})) } \wedge \exk{h+1}{y}\psi(y) \right]  \\
  \end{array}
\end{equation}

\noindent where $[k]=\{1,\dots,k\}$.

More generally, starting from a prenex bijective first-order formula
$\varphi(\barre{t})$ with free variables $\barre{t}$, one eliminates
all quantified variables from the innermost to the outermost one.
This will result in an equivalent Boolean combination of bijective
\textit{atomic} formulas over $\barre{t}$. In the case where $\varphi$ is without free variable (i.e., $\barre{t}$ is
empty), it is easily seen that the elimination process results in a Boolean combination of cardinality statements (note that, of course, $\exists x \varphi(x) \equiv \exk{1}{x} \varphi(x)$). 
\end{proof}

One interesting consequence of Theorem~\ref{TH elimination quantificateur}
is the following result.

\begin{corollary}[Seese~\cite{Seese-96}]~\label{COR MC fobij}
The problem
$\modelchecking{\fobij}$ is decidable in time $O_{\varphi}(|\calS|)$.
\end{corollary}

\begin{proof}
From Theorem~\ref{TH elimination quantificateur}, we know that there exists a Boolean
combination of cardinality statements over the same signature
$\sg$ equivalent to $\Phi$. Given a formula $\exk{k}{x}\Psi(x)$
one can test whether a given $\sigmastructure$ $\calS$ satisfies
$\calS \models \exk{k}{x}\Psi(x)$ in time $O_{\Psi}(|\calS|)$: it
suffices to enumerate all the elements $a$ of the domain, test whether
$(\calS, a) \models \Psi(x)$ in constant time  and count those for
which the answer is positive. If this number is greater than or equal
to $k$ then $ \exk{k}{x}\Psi(x)$ is true in $\calS$. The final
answer for $\Phi$ is given by the boolean combination of the
answers for each cardinality statement.
\end{proof}

\subsubsection{Considerations on an efficient implementation of the algorithm}

Compared to the method of~\cite{Seese-96}, the proofs given in this paper are constructive and easily implementable. But, due to the case of Formula~\ref{EQU cas 2} in Theorem~\ref{TH elimination quantificateur} which leads to the equivalent Formula~\ref{EQU cas 2 bis} the whole process is in $O_{\varphi}(|\calS|)=O(f(|\varphi|).|\calS|)$ for some function $f$ that may be a tower of exponentials.
 It can be shown that it heavily depends on the number of variables and of quantifier alternations of the formula. However, the size of the function $f$ can be substantially reduced in case there are few quantifier alternations.
 
 In what follows, we revisit the method of the proof of Theorem~\ref{TH elimination quantificateur} to prove a slightly different result in a specific case. We focus on formulas with existentially quantified variables only and show that the model-checking problem for such formulas can be efficiently evaluated. A $\fobij$ formula is in $\Sigma_1\!-\!\fobij$  if it is of the form:

\[
\exists \tu y \ \varphi 
\]  

\noindent where $\varphi$ is quantifier-free and in disjunctive normal form (DNF).

\begin{corollary}
 The model-checking problem for $\Sigma_1\!-\!\fobij$ formulas can be evaluated in time $O(|\varphi|^{d}.|\calS|)$  where $d$ is the number of distinct variables of $\varphi$.
\end{corollary}

\begin{proof} 
The result obviously holds for $d=1$. So, assume $d>1$.
For the same reason as in Theorem~\ref{TH elimination quantificateur}, we may consider any formula of the form:

\begin{equation}
\varphi(\barre{x})\equiv \ex y \ (\alpha_1 \wedge \dots \wedge
\alpha_r)
\end{equation}

\noindent where each $\alpha_i$ is a bijective literal~\footnote{In this proof, bijective literals do not involve cardinality statements} 
with variables among $\barre{x}$ and $y$. For sake of completeness here, we consider also terms not containing $y$. Then, $\varphi(\barre{x})$ is of the form:

\begin{equation}
\varphi(\barre{x})\equiv \ex y \ [\psi(y) \wedge y
=_{\eps_1}\t_1(x_{i_1}) \wedge \dots \wedge y
=_{\eps_k}\t_k(x_{i_k}) \wedge \gamma(\tu x)]
\end{equation}

\noindent with the same notation $\eps_{j}$ as in the proof of Theorem~\ref{TH elimination quantificateur} and $\gamma(\tu x)$ involves variables of $\tu x$ only.
Again, if $\eps_{j}=1$, for some $j$, then all the occurences of $y$ are replaced by $\t_j(x_{i_j})$
and $\varphi(\barre{x})$ is equivalent to a conjunction of literals without variable $y$.

Suppose now that $\eps_{j}=-1$ for all $j\leq k$. Let $A=\{a\in D : (\calS,a)\models \psi(y)\}$. Since $\psi(y)$ is quantifier-free, $A$ can be computed in time $O(|\psi|.|\calS|)$. Two cases need to be considered now. If $|A|>k$, since there are at most  $k$ different values $\t_j(x_{j})$ for $j=1,\dots,k$, then the conjunction 
$\ex y  [\psi(y) \wedge y
\neq \t_1(x_{i_1}) \wedge \dots \wedge y
\neq \t_k(x_{i_k})]$ is always true and $\varphi(\tu x)$ is simply equivalent to $\gamma(\tu x)$.
If $|A|\leq k$ let $A= \{a_1,\dots,a_h\}$, with $h\leq k$. Formula $\varphi(\barre{x})$ is replaced by the equivalent formula  below over the richer signature $\sg\cup\{a_1,\dots,a_h\}$:

\[
 \bigvee_{i\leq h} ( \bigwedge_{j\leq
k} a_i\neq \t_j(x_{i_j}) \wedge \gamma(\tu x) )
\] 

In all cases, the formula obtained is also in DNF. 
Time $O(|\varphi|.|\calS|)$ is needed to eliminate variable $y$ and the new formula is of size bounded by $O(k.|\varphi|)$, i.e., less than $O(|\varphi|^2)$.
Elimination of all the $d$ existentially quantified variables except the last one can be pursued from this new formula (without need for a normalisation). In the worst case (where all literals are of the form $
x_i \neq \t_1(x_{j})$), the process will result in a disjunction of less than $|\varphi|^{d-1}$ conjunctions of at most $|\varphi|$ literals.
\end{proof}

\subsection{Constant delay algorithm for first-order queries on bijective structures}

We are now ready to state  the main result of this section.

\begin{theorem}~\label{TH bijective query}
The problem  $\query{\fobij}\in \constantdelaylin$. In particular, from Lemma~\ref{LEM total time}, it can be computed in time $O_{\varphi}(|\calS| + |\varphi(\calS)|)$ and space $O_{\varphi}(|\calS|)$.
\end{theorem}

\begin{definition}
A bijective literal is a bijective atomic formula or its negation.
\end{definition}

Before proving Theorem~\ref{TH bijective query}, we establish the following lemma.

\begin{lemma}~\label{LEM query output}
Let $S$ be a bijective structure and $\Psi$ be a conjunction of bijective literals. Computing  query $\calS \mapsto \Psi(\calS)$ can be done in $\constantdelaylin$.
\end{lemma}

\begin{proof}
The result is proved by induction on $k$ the number of free variables of $\Psi(\tu x)$ where $\tu x = (x_1,\dots,x_k)$. We even assume that $\Psi$ makes use of explicit constants from domain $D$ of $\calS$. 

For the case $k=1$, it is evident that the one variable query $Q = \{a\in D: (\calS, a) \models \Psi(x) \}$ can be evaluated in time $O_{\Psi}(|D|)= O_{\Psi}(|\calS|)$ and hence, by Lemma~\ref{LEM linear time = constant delay}, is in $\constantdelaylin$.

The result is supposed to be true for $k$ ($k\geq 1$) and proved now for $k+1$. Let's consider the query:

\[
 Q = \{(\tu a,b) \in D^{k+1} : \calS \models \Psi(\tu x,y)\}
\]

\noindent where the conjunction of bijective literals $\Psi$ is over variables $\tu x =(x_1,\dots,x_k)$ and $y$. As for Theorem~\ref{TH elimination quantificateur}, two cases need to be distinguished.

\begin{enumerate}

\item~\label{cas 1} $\Psi$ contains at least one literal of the form $\t_1(y) = \t_2(x_{i_0})$, $1\leq i_0 \leq k$, that can also be rephrased as  $y = \tau(x_{i_0})$,

\item~\label{cas 2} $\Psi$ does not contain such a literal.

\end{enumerate}

In the first case, $\Psi$ can rewritten as:

\[ \Psi(\tu x, y) = \Psi_0(\tu x, y) \wedge y = \tau(x_{i_0}).\] 

Query $Q$ is then equivalent to:

\[
Q = \{(\tu a,\tau(a_{i_0})) \in D^{k+1} : (\calS , \tu a ) \models \Psi_0(\tu x,\tau(x_{i_0}))\},
\]

which  is essentially the following $k$ variable query $Q'$: 
\[
Q' = \{\tu a \in D^{k} : (\calS , \tu a ) \models \Psi_0(\tu x,\tau(x_{i_0})\}.
\]

To be precise, $Q=\{(\tu a,\tau(a_{i_0})) : \tu a \in Q'\}$.
By the induction hypothesis, query $Q'$ can be computed by some algorithm $\calA'$ in constant delay. This provides the following constant delay procedure for query $Q$.

\begin{algorithm}
\caption{Evaluating query $Q$}
\begin{algorithmic}[1]
\State \textbf{Input:} $\calS, \Psi$
\State $\precompAlgo{\calA'}$
\State Apply $\enumAlgo{\calA'}$ and for each enumerated tuple $\tu a$, output $(\tu a,\tau(a_{i_0}))$ instead
\end{algorithmic}
\end{algorithm}

\medskip

Case~\ref{cas 2} is a little more complicated. Formula $\Psi$ can be put under the following form:

\[
\Psi \equiv \Psi_1 (\tu x) \wedge \Psi_2 (y) \wedge \bigwedge_{1\leq i \leq r} y \neq \tau_i(x_{j_i})
\]

\noindent with $1\leq j_i\leq k $ for $1\leq i \leq r$. 
By induction hypothesis, the $k$ variable query:

\[
Q_1 = \{\tu a \in D^k : (\calS , \tu a ) \models \Psi_1(\tu x)\}
\]

\noindent can be computed by an algorithm $\calA_1$ on input $\calS$ with constant delay.
For similar reason, the $k$ variable  query $Q_b$ over structure $(\calS,b)$ defined by:

\[
Q_b = \{\tu a \in D^k : (\calS , \tu a, b ) \models \Psi (\tu x, y)\}\}
\]

\noindent can be enumerated by an algorithm using constant delay. Let now $Q_2$ be:

\[
Q_2 = \{ b \in D : (\calS , b)  \models \Psi_2(y)\}.
\]

If $|Q_2|\leq r$ then, by Lemma~\ref{LEM disjoint union}, there exists an algorithm $\calA_0$ which enumerates the disjoint union $\cup_{b\in Q_2} Q_b\times \{b\}$ with constant delay. Note that $\cup_{b\in Q_2} Q_b\times \{b\} = Q$.
From what has been said  Algorithm~\ref{ALGO enum constant delay 1} below correctly computes query $Q$.

\begin{algorithm}[h]
\caption{Evaluating query $Q$}~\label{ALGO enum constant delay 1}
\begin{algorithmic}[1]
\State \textbf{Input:} $\calS, \Psi$
\State Compute $Q_2$ and $|Q_2|$
\If{$|Q_2|\leq r$} run $\calA_0$
\Else 
	\State $\precompAlgo{\calA_1}$~\label{ALGO precomp}
	\For{$\tu a \in \enumAlgo{\calA_1}$}
		\For{$b\in Q_2$}
		\If{$(\calS, \tu a, b)\not\models \bigvee_{1\leq i\leq r} y = \tau_i(x_{j_i})$} Output $(\tu a, b)$
		\EndIf
		\EndFor
	\EndFor
\EndIf	
\end{algorithmic}
\end{algorithm}

\medskip

Up to step~\ref{ALGO precomp} of the algorithm, all can be done in linear time.

 It remains to show that, in the case where $|Q_2|\geq r+1$, the delay between two successive solutions is bounded by some constant. Since $|Q_2|\geq r+1$ and the number of $b\in Q_2$ that verify $(\calS, \tu a, b)\not\models \bigvee_{1\leq i\leq r} y = \tau_i(x_{j_i})$ is bounded by $r$, the algorithm outputs at least one $(\tu a,b)$ for each $\tu a\in Q_1$. More precisely, it outputs $|Q_2|-r$ such tuples. For the same reasons, the maximal delay between two successive outputs is then bounded by $2r$. The same arguments apply for the delay between the last solution and the end of the algorithm. Then, computing $Q$ can be done in constant delay.
\end{proof}

\begin{proofOf}{Theorem~\ref{TH bijective query}}
Let $\calS$ and $\varphi (\tu x)$ be instances of the $\query{\fobij}$ problem. From Theorem~\ref{TH elimination quantificateur}, one can transform $\varphi (\tu x)$ into the following equivalent formula in  disjunctive normal form:

\[
\varphi(\tu x)\equiv \Psi_1(\tu x) \vee \dots \vee \Psi_q(\tu x)
\] 

\noindent where each $\Psi_i$ is a conjunction of bijective literals and for all $i,j$, $1\leq i < j \leq q$ and all bijective structures $\calS$, $\Psi_i(\calS) \cap \Psi_j(\calS) = \emptyset$.  The Theorem immediately follows from Lemma~\ref{LEM disjoint union} since the enumeration problem of each query $\calS \mapsto \Psi_i(\calS)$, $1\leq i \leq q$, belongs to $\constantdelaylin$ by Lemma~\ref{LEM query output}.
\end{proofOf}

\section{Relational structures of  bounded degree}~\label{SEC degre borne}

\subsection{Two equivalent definitions}

Let $\rho = \{R_1,\ldots, R_q\}$ be a relational signature, i.e., a signature made of relational symbols $R_i$ each of arity $a_i$. Recall that $arity(\rho)=max_{1\leq i \leq q} (a_i)=m$.

Let $\calS=\st{D}{\su{R}{q}}$ be a $\rho$-structure. For each $i\leq q$, $R_i\subseteq D^{a_i}$. The \textit{degree} of an element $x$ in $\calS$ is defined as follows:

\[
degree_{\calS}(x) = \sum_{1\leq i \leq q} \sum_{1\leq j \leq a_i} \sharp \{(y_1,\dots,y_{a_i})\in D^{a_i}: \exists j \leq a_i \mbox{ s.t. } x=y_j \mbox{ and } \calS \models R_i(y_1,\dots,y_{a_i})\}.
\]

Intuitively, $degree_{\calS}(x)$ is the total number of tuples of relations $R_i$ to which $x$ belongs to. One  defines the degree of a structure as $degree(\calS) = max_{x\in D} (degree_{\calS}(x))$.

\begin{remark}
In~\cite{Seese-96} a different definition of the degree of a structure is given. It counts, for each  $x$, the number of distinct elements $y\neq x$ adjacent to $x$, i.e., that appear in some tuple with $x$. More precisely, 

\[
degree^1_{\calS}(x) = \sharp \{y : y\neq x \mbox{ and } \exists i \leq q , \tu t \in D^{a_i},  \mbox{ s.t. } \calS \models R_i (\tu t) \mbox{ and } x,y\in \tu t\},
\]

\noindent and $degree^1(\calS) = max_{x\in D} (degree^1_{\calS}(x))$.

Since each tuple containing $x$ contains at most $m-1$ elements different from $x$, it is easily seen that:

\[
degree^1(\calS) \leq (m-1). degree(\calS) \mbox{ where } m=arity(\rho).
\].

Conversely, for  each $x$, if there exist at most $d$ elements $y\in D$ adjacent to $x$ then, the number of distinct tuples involving $x$ and $y$ is bounded by $q.m.d^{m-1}$. Hence,

\[
degree(\calS) \leq q.m.(degree^1(\calS))^{m-1}.
\]

So, the two measures yield the same notion of bounded degree structure.
\end{remark}

We are interested in the complexity of the following query problem for bounded degree structures (which is clearly independent of either measure of degree we choose).
\bigskip

\noindent $\query{\foBoundDeg}$\\
\noindent \textbf{Input:} an integer $d$, a relational signature $\rho$, a $\rho$-structure $\calS$ with $degree(\calS)\leq d$ and a first-order $\rho$-formula  $\varphi(\tu x)$ with $k$ free variables $\tu x = (x_1,\dots,x_k)$\\  
\noindent \textit{Parameter:} $d,\varphi$ \\
\noindent \textbf{Output:} $\varphi(\calS) = \{\tu a \in D^k : (\calS, \tu a) \models \varphi(\tu x)\}$.

\subsection{Interpreting a structure of bounded degree into a bijective  structure}

In this section, we present a natural reduction from $\query{\foBoundDeg}$ to $\query{\fobij}$ which is obtained by interpreting  any structure of bounded degree into a bijective one. 

Let $\calS=\st{D}{\su{R}{q}}$ be a $\rho$-structure of domain $D$, of arity $m=max_{1\leq i \leq q} arity(R_i)$ and of degree bounded by some constant $d$. One associates to $\calS$ a bijective $\sg$-structure $\calS'=\st{D'}{D,\su{T}{q},g,\su{f}{m}}$ of domain $D'$ where $D,\su{T}{q}$ are pairwise disjoints unary relations (i.e. subsets of $D'$) and $g,\su{f}{m}$ are permutations of $D'$. Structure $\calS'$ is precisely defined as follows:

\begin{itemize}

\item $D$ corresponds to the domain of $\calS$.

\item $T_i$ ($1\leq i \leq q$) is a set of elements each representing a tuple of $R_i$ (hence, $card(T_i)=card(R_i)$).

The new domain $D'$ is the disjoint union: $D\cup (D\times\{1,\dots,d\})\cup T_1\cup \dots \cup T_q$. Let us use the following convenient abbreviations: $U = D\cup (D\times\{1,\dots,d\})$ and $T= \bigcup_{1\leq i \leq q} T_i$.

\item $g$ creates a cycle that relates $d$ copies of each element $x$ of the domain. More precisely, for each $x\in D$, it holds $g(x)=(x,1)$, $g((x,i))=(x,i+1)$ for $1 \leq i < d$, and  $g((x,d))=x$. We also set $g(x)=x$ for all other $x$ ($x\in T$).

\item Each $f_i$ is an involutive permutation and essentially represents a projection of $T$ into $D$ as follows.
Let $R_i(x_1,\dots,x_k)$ be true in $\calS$ for some relation $R_i$ of arity $k\leq m$ and some $k$-tuple $(x_1,\dots,x_k)\in D^k$. Suppose $R_i(x_1,\dots,x_k)$ is represented by element $t\in T_i$, then, for each $j\leq k$, set $f_j(t)=(x_j,h)$ and set the reciprocal $f((x_j,h))=t$ if $R(x_1,\dots,x_k)$ is the $h^{th}$ tuple in which $x_j$ appears (with $h\leq d$). The construction is completed by loops $f_j(x)=x$ for all other $x\in D'$.

\end{itemize}       

\begin{figure}[tc]
\begin{center}
\input{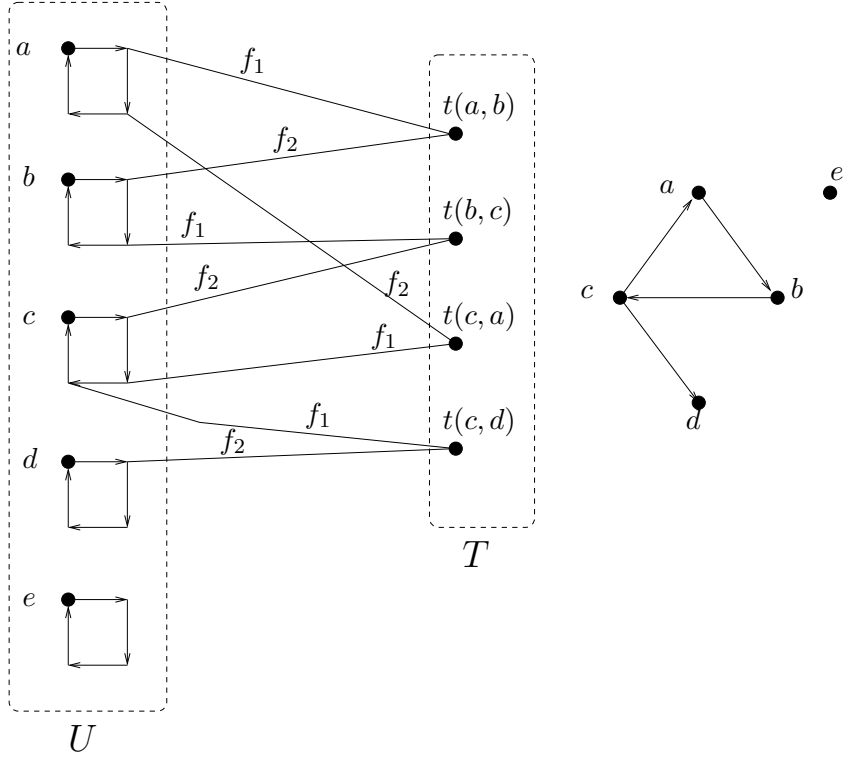}
\end{center}
\caption{Our reduction on an example: the original structure (digraph) of degree $3$ is on the right side of the picture}~\label{ex}
\end{figure}

Figure~\ref{ex} details the reduction on an example.

It is clear that, by construction, $\calS'$ is a bijective structure and that we have the following interpretation Lemma.

\begin{lemma}~\label{LEM interpretation} Let $\theta_i$ be the $\sg$-formula below associated to any symbol $R_i\in\rho$ of arity $k$:

\[
\theta_i(x_1,\dots,x_k)\equiv \exists t (T_i(t) \wedge \bigwedge_{1\leq j\leq k} \bigvee_{1\leq h\leq d} f_j(t)=g^h(x_j)).
\]

Then, for all $(a_1\dots,a_k)\in D^k$:

\[
(\calS, a_1,\dots, a_k) \models R_i(x_1,\dots,x_k) \iff (\calS', a_1,\dots, a_k) \models \theta_i(x_1,\dots,x_k).
\]
\end{lemma} 

To each first-order $\rho$-formula $\varphi(x_1,\dots,x_p)$, one associates the $\sg$-formula $\varphi''(x_1,\dots,x_p)$ obtained by replacing each quantification $\exists v$ (resp. $\forall v$) by the relativized quantification $(\exists v D(v))$ (resp. 
 $(\forall v D(v))$) (that can be written respectively as $\exists v (D(v)\wedge ...)$ and  $\forall v (D(v)\rightarrow ...)$) and by replacing each subformula $R_i(x_1,\dots,x_k)$ by $\theta_i(x_1,\dots,x_k)$.

The following proposition and lemma express that our reduction is correct and linear in $|\calS|$.
Because of Lemma~\ref{LEM interpretation}, Proposition~\ref{PROP interpretation} can be easily proved by induction on formula $\varphi$.

\begin{proposition}[interpretation of $\calS$ into $\calS'$]~\label{PROP interpretation}
For all $(x_1\dots,x_p)\in D^p$:
\[
(\calS, a_1,\dots, a_p) \models \varphi(x_1,\dots,x_p) \iff (\calS', a_1,\dots, a_p) \models \varphi''(x_1,\dots,x_p).
\]

In other words: $\varphi(\calS) = \varphi''(\calS')\cap D^p$. 
Then, setting $\varphi'(x_1,\dots,x_p) \equiv \varphi''(x_1,\dots,x_p)\wedge \bigwedge_{i\leq p}D(x_i)$, it holds: $\varphi(\calS) = \varphi'(\calS')$
\end{proposition}

\begin{lemma}~\label{LEM complexite reduction}
Computing $\calS'$ from $\calS$ can be done in linear time $O_{\rho, d}(|\calS|)$.
\end{lemma}

\begin{proof}
As computing $\calS'$ from $\calS$ is easy, one has only to compare the size of the two structures. The size of $\calS$ is:

\[
|\calS| = \Theta(|D|+\sum_{i=1}^q card(R_i).arity(R_i))= \Theta_{\rho}(|D|+\sum_{i=1}^q card(R_i)).
\] 

For $\calS'$, by construction, it holds that:

\[
|D'|=(d+1).|D| + \sum_{i=1}^q card(R_i)= \Theta_{d,\rho}(|\calS|).
\]

Hence, $|\calS'|=\Theta(m|D'|)=\Theta_{d,\rho}(|\calS|)$.
\end{proof}

\medskip

 We are now ready to state and prove the main result of this section.

\begin{theorem}~\label{TH degre borne}
$\query{\foBoundDeg}$ belongs to $\constantdelaylin$.
\end{theorem}

\begin{proof} 
Let $\calA$ be a constant delay algorithm that computes queries of $\query{\fobij}$. By
using Proposition~\ref{PROP interpretation}, the algorithm below correctly evaluates queries in $\query{\foBoundDeg}$. 

\begin{algorithm*}~\label{ALGO Bounded Degree}
\caption{Evaluating $\query{\foBoundDeg}$}
\begin{algorithmic}[1]
\State \textbf{Input:} $\calS, d, \varphi$
\State Compute the  $\sg$-formula $\varphi'(\tu x)$ associated to $\varphi$ (and $d$)~\label{algo2cpt1}
\State Compute the bijective $\sg$-structure $\calS'$ associated to $\calS$ (and $d$)~\label{algo2cpt2}
\State Run $\calA$ on input $\calS'$, $\varphi'$~\label{algo2cpt3}
\end{algorithmic}
\end{algorithm*}

The cost of instruction~\ref{algo2cpt1} is $O_{\varphi,d}(1)$, that of instruction~\ref{algo2cpt2} is $O_{\varphi,d}(|\calS|)$ (by Lemma~\ref{LEM complexite reduction}) and the precomputation part of algorithm~$\calA$ (included in instruction~\ref{algo2cpt3}) is $O_{\varphi'}(|\calS'|)$ (hence  $O_{\varphi,d}(|\calS|)$) by Theorem~\ref{TH bijective query}.
These steps form a precomputation phase of time complexity $O_{\varphi,d}(|\calS|)$. Finally, the effective enumeration of $\varphi(\calS)=\varphi'(\calS')$ is handled on $\calS', \varphi'$ by $\calA$ and is performed with constant delay. 
\end{proof}

\subsection{Complexity of subgraphs problems}~\label{Complexity of subgraphs problems}

 In this part, we present a simple application of our result to a well-known graph problem.
Given two graphs $G=\st{V}{E}$ and $H=\st{V_H}{E_H}$, $H$ is said to be a \textit{subgraph} (resp. \textit{induced subgraph}) of $G$ if there is a one-to-one function $g$ from $V_H$ to $V$ such that, for all $u,v\in V_H$, $E(g(u),g(v))$ holds if (resp. if and only if) $E_H(u,v)$ holds. 

\bigskip

\noindent \textsc{generate subgraph} (resp. \textsc{generate induced subgraph})\\
\noindent \begin{tabular}{rl}
\textbf{Input:} & \parbox[t]{300 pt}{any graph $H$ and a graph $G$ of degree bounded by $d$}\\
\textbf{Parameter:} & $|H|, d$.\\
\textbf{Output:} & All the subgraphs (resp. induced subgraphs) of $G$ isomorphic to $H$.\\
\end{tabular}

\bigskip

The treewidth of a graph $G$ is the maximal size of a node in a tree decomposition of $G$ (see, for example,~\cite{DowneyF-99}).    
In~\cite{PlehnV-90} it is proved that for graphs $H$ of treewidth at most $w$, testing if a given graph $H$ is an induced subgraph of a  graph $G$ of degree at most $d$ can be done in time $f(|H|,d).|G|^{w+1}$. 
In what follows, we show that there is no reason to focus on graphs of bounded treewidth and that a better bound can be obtained  for \textit{any} graph $H$ (provided $G$ is of bounded degree).
In the result below, we prove that not only the complexity of this decision problem is $f(|H|,d).|G|$ but that generating all the (induced) subgraphs isomorphic to $H$ can be done with constant delay.

 \begin{corollary}
 The problem \textsc{generate subgraph} (resp. \textsc{generate induced subgraph}) belongs to $\constantdelaylin$ 
 \end{corollary}

\begin{proof}
The proof is given for the erate geinduced subgraph problem.
Let $G=\st{V}{E}$ and $H=\st{V_H=\{h_1,\dots,h_k\}}{E_H}$ ($|V_H|=k$) be the two inputs of the problem.
Since $G$ is of maximum degree $d$, we can partition its vertex set $V$ into $d$ sets $V^0,\dots,V^d$ where each $V^{\alpha}$ is the set of vertices of degree $\alpha$.  This can be done in linear time $O(|G|)$. We proceed the same for graph $H$ and obtain the sets $V_H^0, \dots, V_H^d$. In case there exists a vertex in $H$ of degree greater than $d$, it can be concluded immediately that the problem has no solution. Now, let $Q$ be the following formula:

\[
Q (x_1,\dots,x_k) \equiv
\bigwedge_{i<j\leq k} x_i \neq x_j \wedge \bigwedge_{V_H^{\alpha}(h_i)} V_G^{\alpha}(x_i) \wedge \bigwedge_{E_H(h_i,h_j)} E(x_i,x_j).
\] 

Formula $Q$ simply checks that $H$ is a subgraph of $G$ and that each distinguished vertex $x_i$ of $G$ has the same degree as its associated vertex $h_i$ in $H$. Note that formula $Q$ only depends on $H$ and $d$. The result follows now from Theorem~\ref{TH degre borne}.
\end{proof}

\section{Conclusion}

In this paper, we study the complexity of evaluating first-order queries on bounded degree structures and consider this evaluation as a dynamical process, i.e., as an enumeration problem.
Our main contributions are two-fold. First, we define a simple quantifier elimination method suitable for first-order formulas which have to be evaluated against a bijective structure. Second, we define a new complexity class, called $\constantdelaylin$, for enumeration problem which can be seen  as the minimal
robust complexity class for this kind of problems and we prove that our query problem on bounded degree structures belong to this class.

There are several interesting directions for further researches. Among them, the two following series of questions seem worth to be studied:

\begin{itemize}

\item Which "natural" query problems  belong to $\constantdelaylin$ ? More generally, which kind of combinatorial or algorithmic enumeration problems admit constant delay procedures ?

The same questions can be asked for the larger class $\constantdelaypoly$ of constant delay enumeration problems for which polynomial time (instead of linear time) precomputations are allowed.

\item What are the structural properties of the class $\constantdelaylin$ or of the larger  $\constantdelaypoly$ ? Do they have complete problems ? Under which kind of reductions ? Could they be proved to be different from the classes of enumeration problems solvable with linear or polynomial delay ?  

\end{itemize}

\medskip

\textbf{Acknowledgment.} \ We thank Ron Fagin for a very fruitful email exchange  that lead us to define  complexity notions about constant delay computation.

\bibliographystyle{alpha}
\bibliography{../../../biblio/perso,../../../biblio/central}
\nocite{Seese-96,Lindell-04,ComptonH-90,DowneyF-99,Gaifman-82}

\end{document}